\documentclass[preprint,prd,amsmath,amssymb,aps,nofootinbib]{revtex4-1}

\usepackage{graphicx}
\usepackage{siunitx}
\usepackage{bm}
\usepackage{natbib}
\def\mathbi#1{\textbf{\em #1}}

\def\apj{Astrophys.\ J.\ }

\begin{document}
\title{Interpretation of the puzzling gamma-ray spectrum of the Geminga halo}
\author{Kun Fang$^a$}
\author{Xiao-Jun Bi$^{a,b}$}

\affiliation{
$^a$Key Laboratory of Particle Astrophysics, Institute of High Energy
Physics, Chinese Academy of Sciences, Beijing 100049, China \\
$^b$University of Chinese Academy of Sciences, Beijing 100049, China\\
}

\date{\today}

\begin{abstract}
The gamma-ray halo around Geminga is formed owing to the slow diffusion of the electrons released by the Geminga PWN. The latest HAWC and HESS observations exhibit complex features in the TeV gamma-ray spectrum of the Geminga halo. We first show that the new results cannot be interpreted by the commonly used simple model, where a single power-law injection spectrum and an energy index of $\delta=1/3$ for the diffusion coefficient are assumed. We then propose a two-population electron injection model based on the x-ray observations of the Geminga PWN, which consists of a population of freshly accelerated electrons escaping from the PWN through rapid outflows and a population trapped longer inside the PWN before escaping. The two-population model interprets the HAWC and HESS data well, and the goodness of fit improves significantly compared with the single power-law injection model. It also predicts a different energy dependency of the gamma-ray profile from the single power-law model, which could be tested by LHAASO in the coming future. We note that a $\delta$ slightly larger than 1 is needed to fit the HAWC and HESS data consistently. We also discuss the possible improvements by adopting the two-zone diffusion model.
\end{abstract}

\maketitle

\section{Introduction}
\label{sec:intro}
After electrons and positrons\footnote{\textit{Electrons} will denote both electrons and positrons hereafter if not specified.} escape from some middle-aged pulsars (or pulsar wind nebulae, PWNe hereafter), they diffuse very inefficiently in the surrounding interstellar medium (ISM). The diffusion coefficient is several hundred times smaller than the typical value in the Galaxy, and the accumulated electrons generate observable gamma-ray halos through the inverse Compton scattering (ICS) of the background photons, which are known as the pulsar halos \cite{Sudoh:2019lav,Giacinti:2019nbu}. As the gamma-ray morphology traces the spatial distribution of the parent electrons, pulsar halos can be good indicators of electron propagation in localized regions of the Galaxy.

The TeV halo around Geminga is the first discovered and so far best-studied pulsar halo \cite{Abeysekara:2017old}. The unexpected slow-diffusion environment indicated by the pulsar halo revives the discussion on the possibility of Geminga as the source of the cosmic positron excess \cite{Hooper:2017gtd,Fang:2018qco,Profumo:2018fmz,Tang:2018wyr,Shao-Qiang:2018zla,DiMauro:2019yvh,Wang:2021xph}. However, there is no clear conclusion due to the lack of knowledge of the electron injection spectrum and features of the slow-diffusion zone. In the original paper of HAWC \cite{Abeysekara:2017old}, a simple model with a single power-law electron injection spectrum and Kolmogorov's energy dependency of the diffusion coefficient is enough to interpret the observation. We refer to this model as the ``simple model'' hereafter.

The recent gamma-ray spectrum measurements of HAWC and HESS provide more detailed information about the Geminga halo \cite{Zhou:2021dgj,HAWC:2021wia,Mitchell:2021tig}. The latest HAWC spectrum indicates a possible bump feature around 10~TeV \cite{HAWC:2021wia}. Meanwhile, the spectrum unexpectedly climbs again below $\approx3$~TeV. This low-energy feature is also confirmed by the observation of HESS, which measures the gamma-ray spectrum within $\ang{1}$ around Geminga \cite{Mitchell:2021tig}. Given the high angular resolution of HESS, the low-energy gamma-ray component is very likely associated with Geminga. We will show that these new features can no longer be interpreted by the ``simple model.''

In Sec.~\ref{sec:old_model}, we briefly introduce the calculation of the gamma-ray spectrum of the Geminga halo and present the difficulties of interpreting the new spectral features with the ``simple model.'' Considering the electron injection process from the Geminga PWN implied by the x-ray observations, we propose a two population injection model in Sec.~\ref{sec:2pop_model} to better fit the new measurements of the gamma-ray spectrum. In Sec.~\ref{sec:2zone_model}, we further discuss the effect of two-zone diffusion. Finally, we conclude in Sec.~\ref{sec:conclu}.

\section{Difficulties of the ``simple model''}
\label{sec:old_model}
Firstly, we introduce the routine calculation of the gamma-ray spectrum of pulsar halos. We obtain the electron number density around the pulsar by solving the propagation equation and then do the line-of-sight integration to get the electron surface density. The gamma-ray spectrum is derived from the electron surface density through the ICS calculation.

The electron propagation equation can be expressed by
\begin{equation}
  \frac{\partial N(E_e, \mathbi{r}, t)}{\partial t} = \nabla \cdot[D(E_e)\nabla N(E_e, \mathbi{r}, t)] + \frac{\partial[b(E_e)N(E_e, \mathbi{r}, t)]}{\partial E_e} + Q(E_e, \mathbi{r}, t)\,,
 \label{eq:prop}
\end{equation}
where $N$ is the electron number density and $E_e$ is the electron energy. The diffusion coefficient takes the form of $D(E_e)=D_0(E_e/{\rm 100~TeV})^\delta$, where we assume $D_0=3.2\times10^{27}$~cm$^2$~s$^{-1}$ in the present work as measured by HAWC \cite{Abeysekara:2017old}. The second and third terms on the right-hand side are the energy-loss and source terms, respectively. Synchrotron radiation and ICS dominate the energy losses of high-energy electrons. We take the local magnetic field strength (3~$\mu$G, \cite{1996ApJ...458..194M}) for the synchrotron component. We adopt the method introduced in Ref.~\cite{Fang:2020dmi} and the seed photon field given in Ref.~\cite{Abeysekara:2017old} to get the ICS component. 

The source function takes the form of
\begin{equation}
 Q(E_e,\mathbi{r},t)=\left\{
 \begin{aligned}
 & q(E_e)\,\delta(\mathbi{r}-\mathbi{r}_s)\,[(t_s+t_ { \rm
sd})/(t+t_{\rm sd})]^2\,, & t\geq0 \\
 & 0\,, & t<0
 \end{aligned}
 \right.\,,
 \label{eq:src}
\end{equation}
where $q(E_e)$ is the current electron injection spectrum, $\mathbi{r}_s$ and $t_s$ are the position and age of Geminga, respectively, and $t_{\rm sd}$ is the pulsar spin-down time scale, which is set to be 10~kyr. The time profile of the source function is assumed to follow the pulsar spin-down luminosity, and $t=0$ corresponds to the birth time of Geminga.

The solution of Eq.~(\ref{eq:prop}) can be expressed by
\begin{equation}
 N(E_e, \mathbi{r}, t) = \int_{R^3} d^3\mathbi{r}_0\int_{t_{\rm ini}}^{t}dt_0\, \frac{b(E_e^\star)}{b(E_e)}\frac{1}{(\pi\lambda^2)^{3/2}}{\rm exp}\left[-\frac{(\mathbi{r}-\mathbi{r}_0)^2}{\lambda^2}\right]\,Q(E_e^\star, \mathbi{r}_0, t_0)\,,
 \label{eq:solution}
\end{equation}
where
\begin{equation}
 E_e^\star\approx \frac{E_e}{[1-b_0E_e(t-t_0)]}\,,\quad\lambda^2=4\int_{E_e}^{E_e^\star}\frac{D(E'_e)}{b(E'_e)}dE'_e\,,
 \label{eq:cooling}
\end{equation}
and $t_{\rm ini}={\rm max}\{t-1/(b_0E_e), 0\}$. We integrate $N$ over the line of sight from Earth to the vicinity of the pulsar and get the electron surface density $S_e(\theta)=\int_0^{\infty}N(l_\theta)dl_\theta$, where $\theta$ is the angle observed away from the pulsar, $l_\theta$ is the length in that direction, and $N(l_\theta)$ is the electron number density at a distance of $\sqrt{d^2+l_\theta^2-2dl_\theta\cos\theta}$ from the pulsar, where $d$ is the distance between the pulsar and Earth. The gamma-ray surface brightness $S_\gamma(\theta,E_\gamma)$ is derived from the electron number density and the standard calculation of ICS \cite{Blumenthal:1970gc}. Finally, we can get the gamma-ray spectrum $F_{\theta_0}(E_\gamma)$ within an arbitrary angular radius $\theta_0$ around the pulsar by $F_{\theta_0}(E_\gamma)=\int_{\ang{0}}^{\theta_0}S_\gamma(\theta,E_\gamma)2\pi\theta d\theta$.

\begin{figure}[t]
\centering
\includegraphics[width=0.48\textwidth]{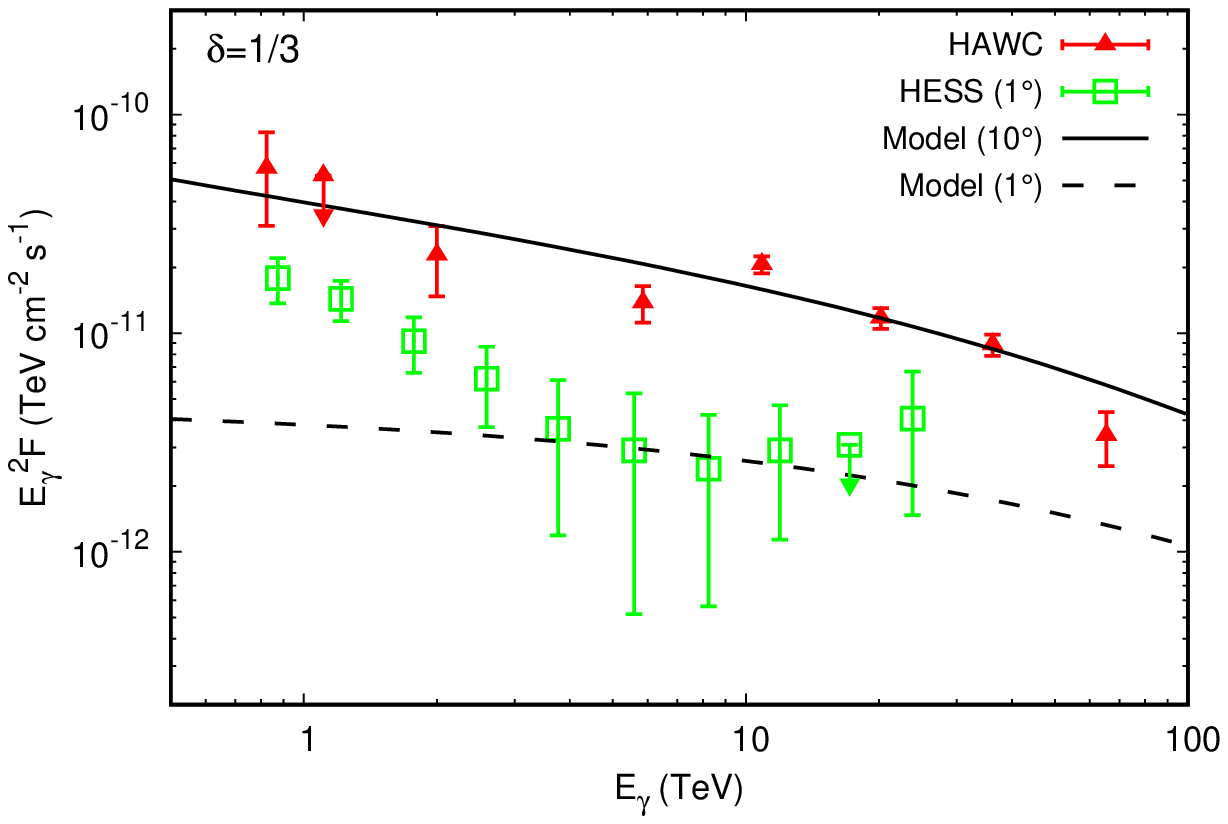}
\includegraphics[width=0.48\textwidth]{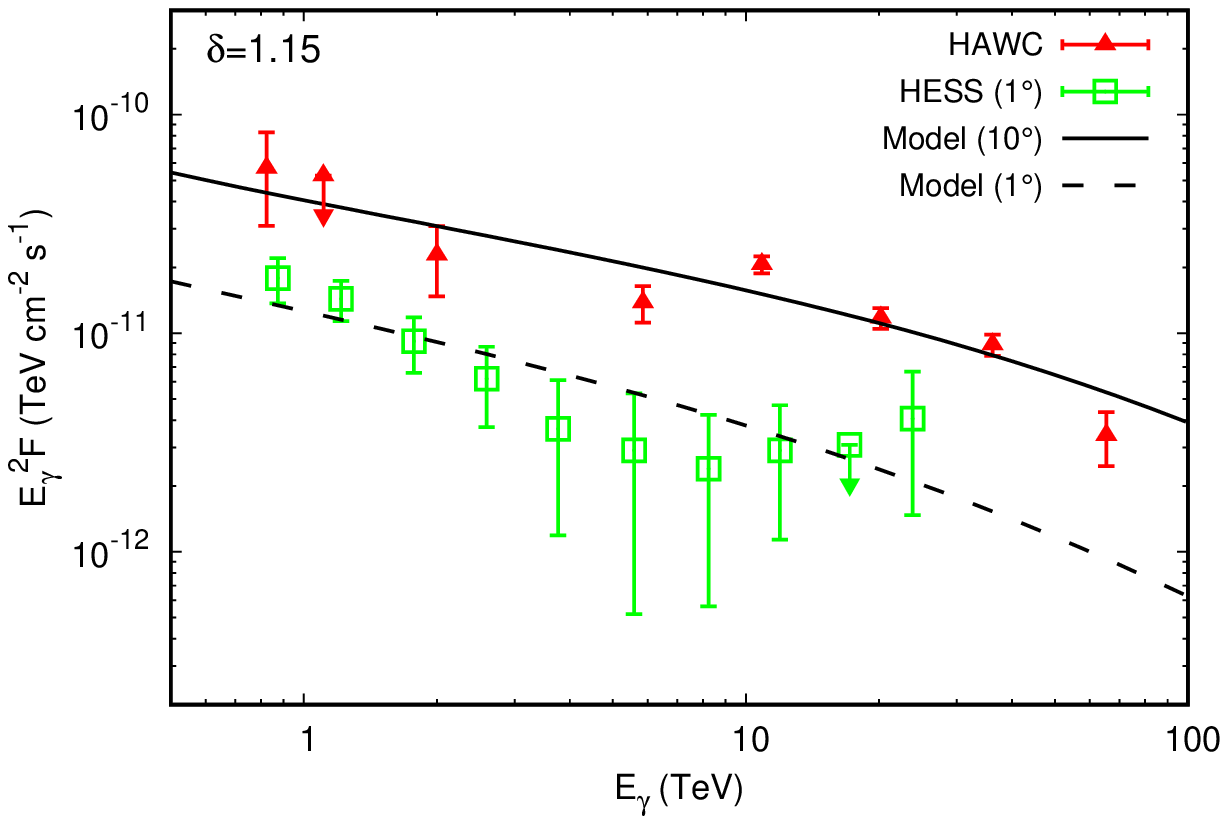}
\caption{Left: the gamma-ray spectra of the Geminga halo given by the "simple model" (a single power-law injection spectrum and Kolmogorov's energy index of the diffusion coefficient for electrons), compared with the HAWC \cite{HAWC:2021wia} and HESS \cite{Mitchell:2021tig} measurements. Right: the same as the left, while the energy index of the diffusion coefficient, $\delta$, is set as a free parameter to fit the HAWC and HESS data. The best-fit $\delta$ is 1.15.}
\label{fig:old_model}
\end{figure}

For the ``simple model,'' the injection spectrum takes the form of $q(E_e)\propto E_e^{-p}$. The normalization of the injection spectrum can be obtained by the relation of $\int_{\rm 1GeV}^{\infty} q(E_e)E_edE_e=\eta L$, where $L$ is the current pulsar spin-down luminosity, and $\eta$ is the conversion efficiency from the spin-down energy to the electron energy. The pulsar age, distance, and spin-down luminosity are 342~kyr, 250~pc, and $3.25\times10^{34}$~erg~s$^{-1}$, which can be found in the Australia Telescope National Facility catalog \cite{Manchester:2004bp}. We set $p$ and $\eta$ as the free parameters in the following fit. The energy index of the diffusion coefficient, $\delta$, is set to be $1/3$ as predicted by Kolmogorov's theory.

We fit the ``simple model'' to the HAWC gamma-ray spectrum by the minimum chi-square method and then compare the best-fit model with the HESS spectrum. The best-fit parameters are $p=2.39$ and $\eta=5.46$\footnote{The required conversion efficiency is larger than 1, while it is not a serious problem for a power-law injection spectrum with $p>2.0$. In this case, the energy of the electron spectrum is concentrated at GeV energy range, where the injection spectrum is not constrained. A low-energy spectral break or cutoff can avoid a too large $\eta$.}, respectively. The result is presented in the left panel of Fig.~\ref{fig:old_model}. There are two serious problems in the result. Firstly, the best-fit model to the HAWC data fits poorly to the HESS spectrum, especially for the low-energy data. It means that the HAWC and HESS results cannot be consistently interpreted with the ``simple model.'' The gamma-ray flux ratio $F_{\ang{1}}/F_{\ang{10}}$ predicted by the ``simple model'' increases with energy, while the measurements show a contrary tendency. Secondly, the fit to the HAWC spectrum alone is also poor---the reduced $\chi^2$ is $\approx3.5$. A single power-law injection spectrum cannot reproduce the possible spectral bump and high-energy cutoff features. 

The small $F_{\ang{1}}/F_{\ang{10}}$ below $\approx3$~TeV indicates that the spatial distribution of low-energy electrons could be more concentrated than expected. A $\delta$ larger than the Kolmogorov value is the most straightforward solution to this problem. We add $\delta$ as a free parameter and simultaneously fit the HAWC and HESS spectrum. The best-fit parameters are $p=2.39$, $\eta=5.21$, and $\delta=1.15$. The required $\delta$ is significantly larger than $1/3$ ($1/2$) as predicted by Kolmogorov's (Kraichnan's) theory and close to the value of Bohm diffusion. As shown in the right panel of Fig.~\ref{fig:old_model}, a consistent interpretation of the two spectra seems to be achieved. However, the reduced $\chi^2$ of the fit is 1.85, and the model is rejected by the goodness-of-fit test at a confidence level of $97\%$. The main reason is still the poor fit to the HAWC spectrum.

\section{A two-population injection model}
\label{sec:2pop_model}

\subsection{Description of the model}
\label{subsec:descri}
Electron-positron pairs produced by pulsars are accelerated to very high energies in the PWNe and then released into the ISM to generate pulsar halos. Thus, observations of the non-thermal radiation of PWNe are helpful for inferring the electron injection spectrum. The Geminga PWN has been observed in x-rays by XMM-Newton and Chandra telescopes with high spatial resolution \cite{2003Sci...301.1345C,2006ApJ...643.1146P,2010ApJ...715...66P,Posselt:2016lot}. The most distinct feature of the PWN is the three-tail structure, consisting of two lateral tails and an axial tail. The latest Chandra observation indicates that the lateral tails, which directly connect to the pulsar, are more likely to be interpreted by outflows induced by polar jets of the pulsar than the limb-brightened shell of the PWN \cite{Posselt:2016lot,Hui:2017xbw}. The axial tail could be interpreted as a crushed torus \cite{Posselt:2016lot} or the bow-shock nebula \cite{Hui:2017xbw}.

\begin{figure}[t]
\centering
\includegraphics[width=0.48\textwidth]{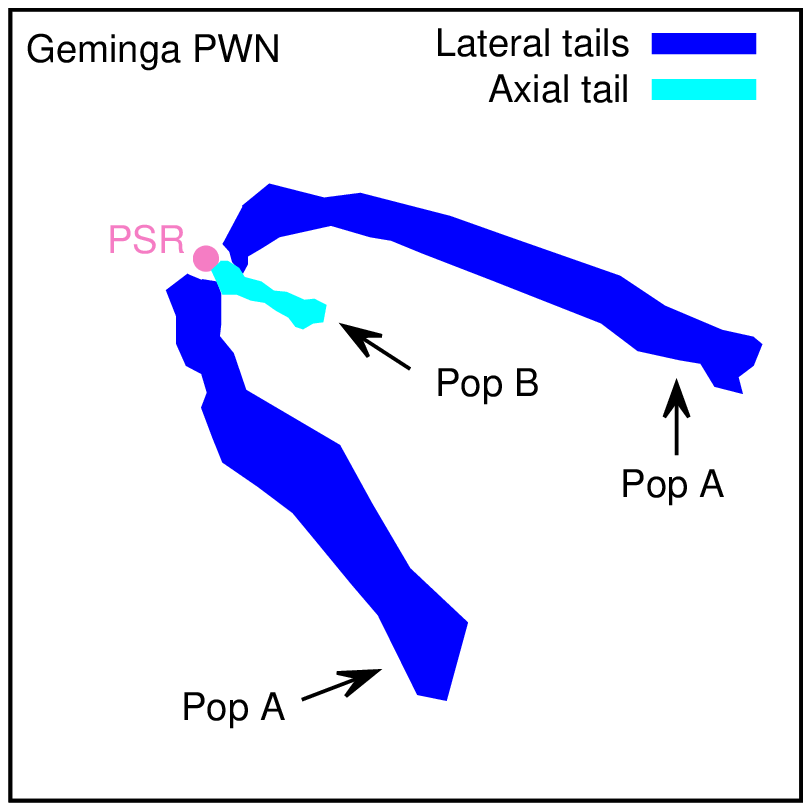}
\includegraphics[width=0.48\textwidth]{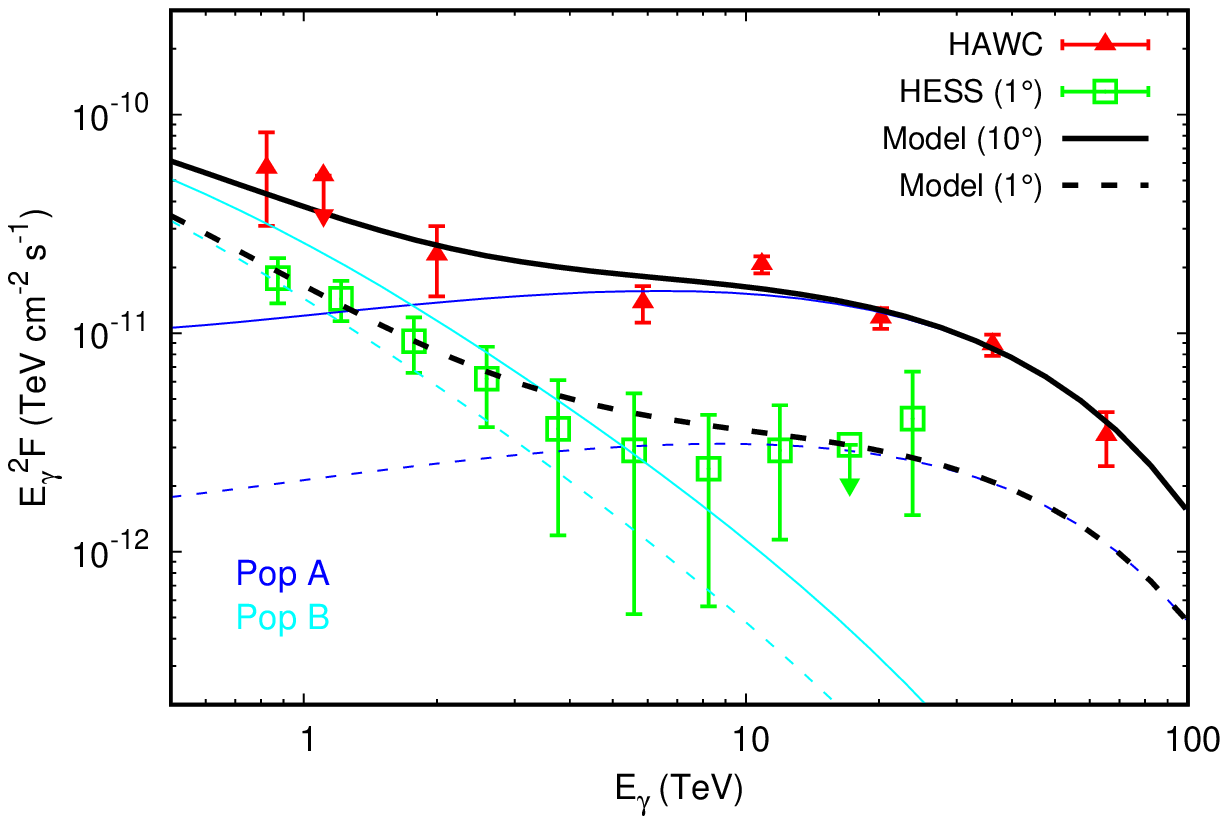}
\caption{Left: sketch of the two-population injection model based on the morphology of the Geminga x-ray PWN \cite{Posselt:2016lot}. Right: best-fit gamma-ray spectra to the HAWC and HESS data with the two-population injection model. For each population, the spectrum within $\ang{10}$ ($\ang{1}$) field around the pulsar is drawn with solid (dotted) line.}
\label{fig:2pop_model}
\end{figure}

According to the image and spectral measurements of the Geminga PWN, we propose a two-population electron injection model and illustrate it with the sketch in the left of Fig.~\ref{fig:2pop_model}. Obviously, electrons can be injected through the lateral outflows, and we name this population Pop A. As there is no evidence of synchrotron cooling in the lateral tails, the velocity of the outflows must be larger than $\sim1000$~km~s$^{-1}$ \cite{Posselt:2016lot}. Moreover, Ref.~\cite{Hui:2017xbw} suggests that the outflows are mild relativistic. The observed brightness difference between the northern and southern tails could be interpreted by Doppler boosting of the high-speed flows. All these indicate that Pop A is fresh electrons efficiently escaping from the acceleration sites, which is consistent with the very hard x-ray spectrum of the lateral tails. Besides, the energy of the parent electrons of the gamma-ray halo can reach $\sim100$~TeV. The synchrotron lifetime of 100~TeV electrons is $\approx200$~yr considering the $20$~$\mu$G magnetic field in the Geminga PWN \cite{Posselt:2016lot}, which means that part of the high-energy electrons must escape from the PWN within 200~yr after being accelerated. Therefore, Pop A is expected to explain the high-energy part of the halo spectrum.

We assume a super-exponentially-cutoff power law for the injection spectrum of Pop A:
\begin{equation}
 q_A(E_e)\propto E_e^{-p}\,{\rm exp}\left[-\left(\frac{E_e}{E_c}\right)^2\right]\,.
 \label{eq:qa}
\end{equation}
The electron spectral indices of the northern and southern tails of the x-ray PWN are 0.67 and 1.04, respectively \cite{Posselt:2016lot}, and we take the average value of 0.85 for $p$. The cutoff term in Eq.~(\ref{eq:qa}) describes the acceleration limit of the PWN, the form of which is suggested by Ref.~\cite{Zirakashvili:2006pv}. The cutoff energy $E_c$ is set as a free parameter. Another free parameter of Pop A is the conversion efficiency $\eta_A$, which corresponds to the normalization of the injection spectrum.

Meanwhile, electrons may also escape from the axial tail in a less efficient way, and we name this population Pop B. The x-ray observations show a tendency of spectral softening along the axial tail \cite{Posselt:2016lot}, which may be due to the synchrotron cooling of electrons. This implies that Pop B may be trapped longer in the PWN than Pop A before being injected into the ISM. Suppose the trapping time is $\sim1000$~yr, a spectral break at $\sim10$~TeV is expected for the electron injection spectrum. We expect that Pop B dominates the low-energy part of the gamma-ray spectrum. 

Compared with Eq.~(\ref{eq:qa}), a spectral break is added to the injection spectrum of Pop B:
\begin{equation}
 q_B(E_e)\propto E_e^{-p}\,\left[1+\left(\frac{E_e}{E_b}\right)^s\right]^{\Delta p/s}\,{\rm exp}\left[-\left(\frac{E_e}{E_c}\right)^2\right]\,,
 \label{eq:qb}
\end{equation}
where the break energy $E_b$ is set as a free parameter. The smooth parameter $s$ has little effect on the result and is set to be $5$. As the injection spectrum of Pop A is the spectrum of freshly accelerated electrons, we assume that Pop B shares the same $p$ and $E_c$ with Pop A. The energy range of x-ray observations is 0.3-8~keV, corresponding to the electron energy of $\approx25-130$~TeV. As this energy range is between $E_b$ and $E_c$ (we will show below that $E_c$ is larger than 130~TeV as required by the fit), we may estimate $-p+\Delta p$ by the x-ray spectral index of the axial tail. We adopt the index of the region relatively far away from the pulsar (the A2$+$A3 region\footnote{In Ref.~\cite{Posselt:2016lot}, the A4 region is the farthest axial region to the pulsar, while the spectral measurement of the A4 region may be contaminated by a known star.} shown in Ref.~\cite{Posselt:2016lot}) and get $\Delta p=-2.85$. This electron spectrum is steeper than that predicted by a constant-injection-cooling scenario and flatter than that of a pure cooling scenario, which may be ascribed to the complex particle transport in the bow-shock nebula. In addition, the conversion efficiency for Pop B, $\eta_B$, is set to be a free parameter. 

\subsection{Fitting result and discussion}
\label{subsec:result}
In summary, the free parameters of the two-population model are $\eta_A$, $\eta_B$, $E_c$, $E_b$, and $\delta$. We fit the two-population model to both the HAWC and HESS data and present the fitting result in the right panel of Fig.~\ref{fig:2pop_model}. The reduced $\chi^2$ of the best-fit result is 1.10, which improves significantly compared with the one-population models in Sec.~\ref{sec:old_model}. The high-energy cutoff term of Pop A can well reproduce the high-energy HAWC data, and the hard power-law term of Pop A gives a better fit to the spectral flattening just below $\approx10$~TeV of the HAWC data. The gamma-ray spectrum generated by Pop B dominates the low-energy range and account for the spectral climb of the HAWC and HESS data below $\approx5$~TeV.

The best-fit parameters are $\eta_A=0.0725$, $\eta_B=0.325$, $E_c=143$~TeV, $E_b=3.93$~TeV, and $\delta=1.25$. The total conversion efficiency from the pulsar spin-down energy to the injected electron energy is $\approx40\%$, which is in a reasonable range. The observations of the x-ray PWN indicate that the Geminga PWN can accelerate electrons up to $\sim100$~TeV. The best-fit $E_c$ is larger than 100~TeV, consistent with the x-ray observations. The best-fit breaking energy of Pop B corresponds to a synchrotron cooling time of $\approx5000$~yr, indicating that Pop B may be trapped inside the PWN for thousands of years before escaping. The best-fit $\delta$ is similar to that obtained with the single power-law injection model in Sec.~\ref{sec:old_model}. 

As the electron injection process of Pop A is evident, we may also give a rough estimate of the injection rate of Pop A based on the x-ray observations. The brighter part of the southern tail measured by Ref.~\cite{2010ApJ...715...66P} is used for the estimate. The unabsorbed luminosity in the $0.3-8$~keV band of this region is $\mathcal{L}=1.96\times10^{29}$~erg~s$^{-1}$. The outflow velocity is assumed as $v\sim c$. As the outflow is believed to be bent by the ram pressure, only the length in the initial jet direction of this region is useful for estimating the injection rate, which is $l\approx75"\times\pi/\ang{180}\times250~{\rm pc}=0.09$~pc. We take the approximation provided by Ref.~\cite{Longair:1994wu} for the calculation of synchrotron emission, which can be expressed by
\begin{equation}
 \mathcal{L}=\int_{E_{e,1}}^{E_{e,2}}b(E_e)\mathcal{Q}(E_e)dE_e\,,
 \label{eq:sync}
\end{equation}
where $E_{e,1}$ and $E_{e,2}$ are the electron energies corresponding to synchrotron peak frequencies of $0.3$ and $8$~keV, respectively, and $b$ is the synchrotron energy-loss rate. The electron energy spectrum $\mathcal{Q}$ has the same form as Eq.~(\ref{eq:qa}), while the normalization is determined by Eq.~(\ref{eq:sync}). The total electron energy in this bright region is $\mathcal{E}=\int\mathcal{Q}(E_e)E_edE_e$. Finally, the total injection rate of the lateral outflows is estimated by $2\mathcal{E}v/l$, and the conversion efficiency from the pulsar spin-down luminosity to Pop A is $2\mathcal{E}v/l/L\approx2\%$, where $L$ is the current spin-down luminosity of Geminga. This efficiency is about three times smaller than the best-fit $\eta_A$. However, the value of $\mathcal{L}$ is based on the assumption of isotropic emission. Considering the large angle between the outflows and the line of sight, the real $\mathcal{L}$ could be significantly larger, and the conversion efficiency estimated by the x-ray observation could be in better agreement with the fitting result.

\begin{figure}[t]
\centering
\includegraphics[width=0.5\textwidth]{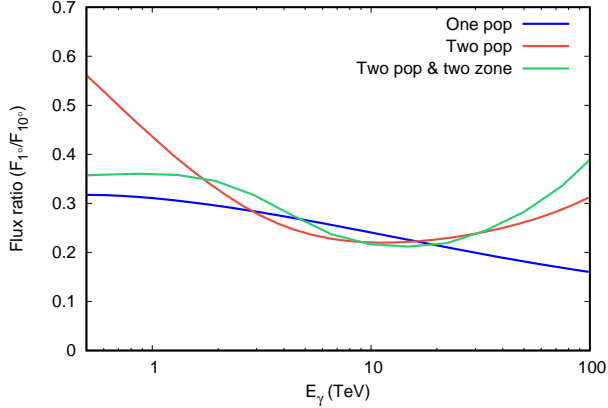}
\caption{Flux ratio of $F_{\ang{1}}/F_{\ang{10}}$ predicted by different models. The blue, red, and green lines correspond to the one-population single power-law injection model in the right of Fig.~\ref{fig:old_model}, the two-population injection model in Fig.~\ref{fig:2pop_model}, and the two-population injection model with two-diffusion assumption in Fig.~\ref{fig:2z_model}, respectively.}  
\label{fig:ratio}
\end{figure}

Apart from the improvement of the goodness of fit, the two-population model also predicts a different energy dependency of the gamma-ray flux ratio $F_{\ang{1}}/F_{\ang{10}}$ from the single power-law model. As shown in Fig.~\ref{fig:ratio}, the single power-law model in the right of Fig.~\ref{fig:old_model} predicts a monotonically decreasing flux ratio, while the flux ratio corresponds to the two-population model in Fig.~\ref{fig:2pop_model} decreases with energy in the low-energy range and increases again above $\sim10$~TeV. The difference is due to the different injection spectra of these two models. In the case of continuous injection, the injection spectrum may significantly affect the spatial distribution of electrons. We give a qualitative explanation below. 

Considering a constant injection case, the electron distribution can be expressed by 
\begin{equation}
 N(E_e,r)=\int_{E_e}^\infty dE_0\,\frac{b(E_0)}{b(E_e)}\,\frac{1}{(\pi\lambda^2)^{3/2}}\,{\rm exp}\left(-\frac{r^2}{\lambda^2}\right)\,Q(E_0)\,.
\end{equation}
It can be seen that $N(E_e,r)$ is a superposition of electron distributions with different initial energies $E_0$, and $Q(E_0)$ is the weight of the superposition. For $\delta>1$, electrons with larger energies have more extended distributions, and the gamma-ray flux ratio $F_{\ang{1}}/F_{\ang{10}}$ tends to decrease with energy. However, the two-population model assumes a cutoff term for the injection spectrum. The number of electrons above the cutoff energy decreases sharply, which means that the spatially extended component of the electron distribution is seriously reduced. Thus, electron distributions with higher energies are less extended, and $F_{\ang{1}}/F_{\ang{10}}$ increases with energy.

The energy dependency of the gamma-ray flux ratio can be a good criterion for models. Due to the large uncertainties of the HESS spectrum above $\approx5$~TeV, the energy dependency in high energies is not well constrained. The LHAASO experiment \cite{Bai:2019khm} is expected to provide morphology measurements for gamma-ray pulsar halos in a wide energy range in the coming future, which may give clear judgments to different models.

\section{Two-zone diffusion}
\label{sec:2zone_model}
In the above calculations, the electron diffusion is assumed to be homogeneous. However, the slow-diffusion process around Geminga should not be typical in the Galaxy. Considering the possible origins of the slow-diffusion environment, the two-zone diffusion model may be a more reasonable assumption (e.g., Ref.~\cite{Fang:2018qco}). The slow-diffusion zone could be ascribed to the streaming instability induced by the electrons escaping from Geminga \cite{Evoli:2018aza,Mukhopadhyay:2021dyh}. It could also be generated by an external source, such as the parent SNR of Geminga, which may provide enough energy for the slow-diffusion environment \cite{Fang:2019ayz}. Either the interpretation suggests a slow-diffusion zone with a size of $\sim50$~pc. Furthermore, the multiwavelength gamma-ray spectrum of another pulsar halo LHAASO J0622$+$3755 also indicates that the slow-diffusion zone may not be larger than $\sim50$~pc \cite{Fang:2021qon}. Below we discuss the effects of two-zone diffusion, based on the two-population injection model in Sec.~\ref{sec:2pop_model}.

Firstly, the two-zone diffusion model can fit the observations with a $\delta$ smaller than the one-zone diffusion model required. When electrons escape from the slow-diffusion zone, they rapidly spread in the ISM due to the much larger diffusion coefficient outside. In this case, part of electrons is still accumulated nearby the pulsar (e.g., within $\ang{1}$ field around the pulsar), while the electrons injected earlier may escape very far away from the pulsar (e.g., out of $\ang{10}$ field around the pulsar). This feature is clearly illustrated by Fig.~1 of Ref.~\cite{Fang:2018qco}. Thus, the two-zone diffusion model predicts a larger $F_{\ang{1}}/F_{\ang{10}}$ than the one-zone diffusion case for the same $\delta$. For $\delta<1$, low-energy electrons have larger propagation scales and can escape from the slow-diffusion zone more easily. It means that the increase of $F_{\ang{1}}/F_{\ang{10}}$ is more significant for the low-energy range of the gamma-ray spectrum, which may explain the large $F_{\ang{1}}/F_{\ang{10}}$ indicated by the HAWC and HESS data $\lesssim5$~TeV.

\begin{figure}[t]
\centering
\includegraphics[width=0.48\textwidth]{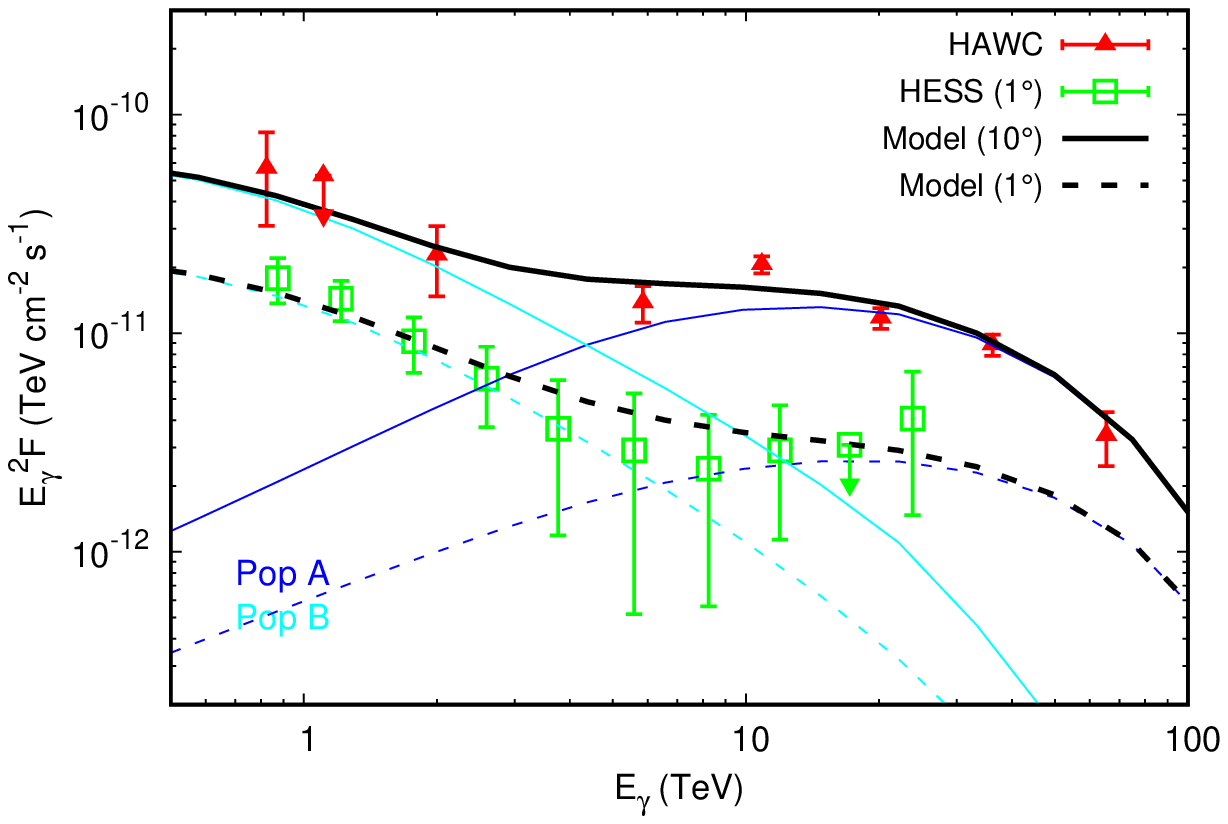}
\includegraphics[width=0.48\textwidth]{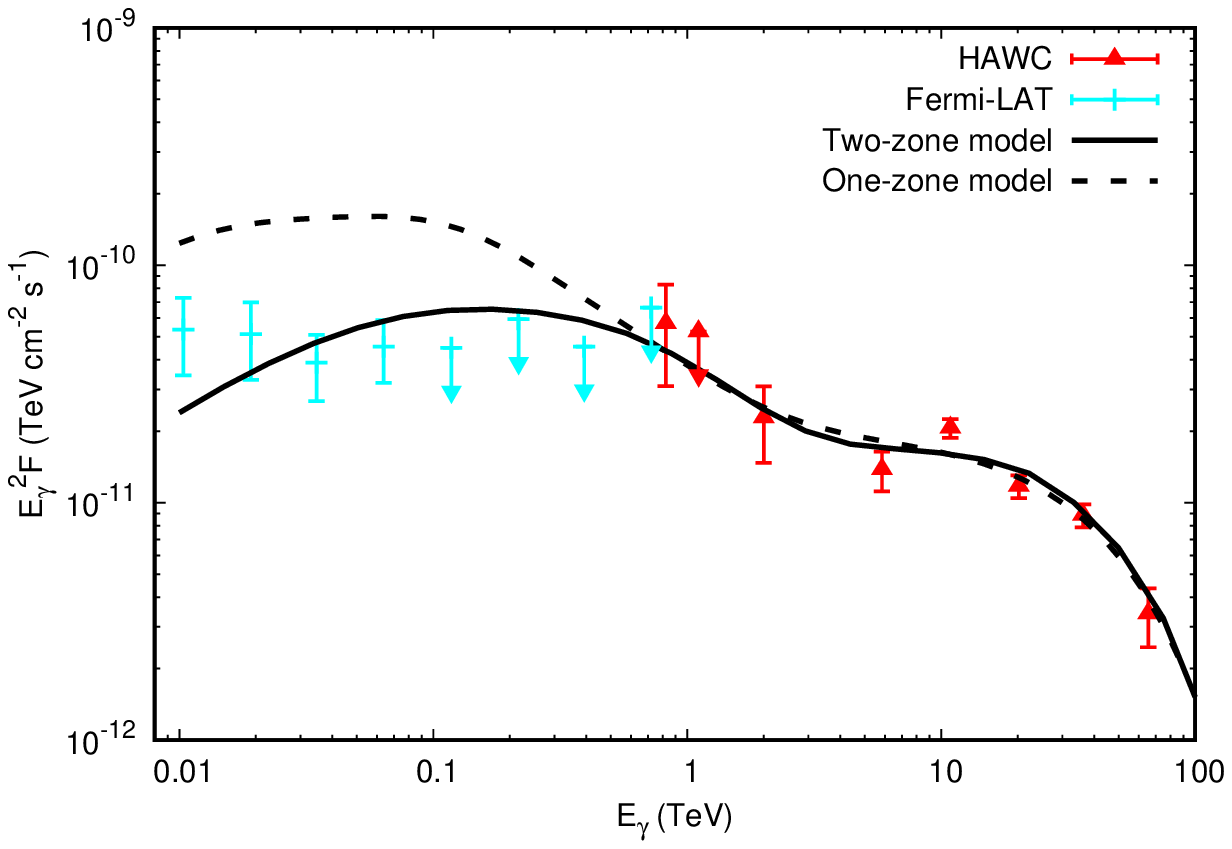}
\caption{Left: spectra calculated under the two-zone diffusion assumption, based on the two-population injection model in Fig.~\ref{fig:2pop_model}. Right: wide band gamma-ray spectra extrapolated from the two-zone model in the left and the one-zone model in Fig.~\ref{fig:2pop_model}, compared with the Fermi-LAT data obtained by Ref.~\cite{DiMauro:2019yvh}.}
\label{fig:2z_model}
\end{figure}

We assume $\delta=0.5$ that is predicted by Kraichnan's theory and adopt the two-zone diffusion model to explain the HAWC and HESS data. The diffusion coefficient takes the form of 
\begin{equation}
 D(E_e, \mathbi{r})=\left\{
 \begin{aligned}
  & D_1(E_e)\,,\quad |\mathbi{r}-\mathbi{r}_s|<r_\star \\
  & D_2(E_e)\,,\quad |\mathbi{r}-\mathbi{r}_s|\geq r_\star \\
 \end{aligned}
 \right.,
 \label{eq:2zone}
\end{equation}
where $D_1$ is the slow-diffusion coefficient used above, $D_2$ is the typical diffusion coefficient of the Galaxy \cite{Yuan:2017ozr}, and $r_\star$ is the size of the slow-diffusion zone. The gamma-ray profile measured by HAWC indicates that $r_\star$ should not be smaller than $\approx25$~pc for high-energy electrons, while we find that a smaller $r_\star$ is needed to interpret the large $F_{\ang{1}}/F_{\ang{10}}$ in the low-energy range. Therefore, we assume $r_\star=30$~pc for $E_e\geq30$~TeV, and $r_\star=15$~pc for $E_e<30$~TeV. In fact, an energy-dependent $r_\star$ is also suggested by the self-excited scenario of slow diffusion \cite{Evoli:2018aza}. We show the comparison between the two-zone model and the observations in the left panel of Fig.~\ref{fig:2z_model}. The low-energy data are well explained as expected. Most of the free parameters used here are similar to the model in Fig.~\ref{fig:2pop_model}, while a larger conversion efficiency of 1.25 is required for Pop B due to the smaller $r_\star$ in low energies. We also show the energy dependency of the flux ratio of this model in Fig.~\ref{fig:ratio}, which has a similar tendency with the model in Fig.~\ref{fig:2pop_model}.

Besides, the two-zone diffusion model may have a better agreement with the GeV observation of the Geminga halo. The GeV Geminga halo is expected to be very extended, and the spectral measurement is challenging due to the large uncertainty of the gamma-ray background \cite{Shao-Qiang:2018zla,DiMauro:2019yvh}. In the right panel of Fig.~\ref{fig:2z_model}, we give a comparison between the models in this work and the Fermi-LAT spectrum of the Geminga halo obtained by Ref.~\cite{DiMauro:2019yvh}. The GeV spectrum is significantly affected by the history of the diffusion pattern. For the self-generated scenario of slow diffusion, the diffusion coefficient began to decrease to the current level $\sim10^5$~yr after the birth of Geminga \cite{Mukhopadhyay:2021dyh}. For the alternative scenario proposed by Ref.~\cite{Fang:2019ayz}, a time scale of $\sim10^5$~yr may also be needed for Geminga to approach the post-shock region, where a slow-diffusion environment is expected. Thus, we assume a $10^5$~yr delay for the emergence of the slow-diffusion zone. As shown in Fig.~\ref{fig:2z_model}, the GeV spectrum extrapolated from the one-zone diffusion model in Fig.~\ref{fig:2pop_model} is several times higher than the Fermi-LAT data, while the GeV fluxes predicted by the two-zone diffusion model is significantly suppressed and consistent with the Fermi-LAT spectrum. It should be emphasized that the above comparison is qualitative. The calculation of the GeV spectrum is affected by the low-energy electron injection spectrum, the time evolution and energy dependency of $r_\star$, and the energy dependency of the diffusion coefficient, all of which are currently not constrained. 

\section{Conclusion}
\label{sec:conclu}
In this work, we first show that the latest HAWC and HESS gamma-ray spectra of the Geminga halo can no longer be interpreted by the ``simple model'', where a single power-law injection spectrum and Kolmogorov's energy index of the diffusion coefficient ($\delta=1/3$) are assumed. Although a larger $\delta$ may account for the unexpected large flux ratio between the HESS and HAWC data below $\approx5$~TeV, the single power-law assumption for electron injection is disfavored by the goodness-of-fit test. The reason is that the gamma-ray spectrum derived from this simple assumption cannot reproduce the complex spectral features of the measurements.

The Geminga PWN is the source of the electrons that light up the Geminga halo, and the x-ray PWN and the TeV gamma-ray halo are generated by electrons with almost the same energy range. Thus, it is meaningful to estimate the electron injection spectrum by the x-ray observations. We propose a two-population injection model based on the image and spectral measurements of the Geminga x-ray PWN. One population (Pop A) is the freshly accelerated electrons that escape from the PWN through rapid outflows, corresponding to the lateral tails of the x-ray PWN. The spectrum of Pop A consists of a hard power-law term and a high-energy cutoff term that describes the acceleration limit. The other population (Pop B) is the electrons trapped longer in the PWN before escaping, which corresponds to the axial tail of the x-ray PWN. A spectral break is further assumed in the injection spectrum of Pop B due to the synchrotron cooling inside the PWN.

The two-population injection model can consistently fit the HAWC and HESS spectra with a reduced $\chi^2$ of 1.10, compared with $\chi^2=1.85$ for the single power-law injection model. The high-energy HAWC spectrum is reproduced by the cutoff term of the injection spectrum, and the HESS and low-energy HAWC spectral features are well interpreted by the superposition of Pop A and B. The required injection rate of Pop A is consistent with that roughly derived from the x-ray observation, which further supports this model. Intriguingly, a $\delta$ slightly larger than 1 is needed to fit the data. It is larger than that predicted by Kolmogorov's or Kraichnan's theory and closer to the value of Bohm diffusion. The two-population model also predicts a different energy dependency of the gamma-ray profile from the single power-law model, which will be tested by the energy-dependent morphological measurements of LHAASO.

We also discuss the effect of two-zone diffusion on electron propagation, which could be a more reasonable scenario considering the possible origins of slow diffusion. Compared with the one-zone diffusion case, a smaller $\delta$ is needed to interpret the HAWC and HESS data. Besides, the GeV spectrum predicted by the two-zone diffusion model may have better consistency with the Fermi-LAT observation of the Geminga halo, although the uncertainties of the data analysis and spectrum calculation in the GeV band are both large at present.

\begin{acknowledgments}
We thank P.~Yin, X.~Chen, and S.~Xi for the helpful discussions. This work is supported by the National Natural Science Foundation of China under Grants 
No. 12175248 and No. 12105292.
\end{acknowledgments}

\bibliography{references}

\end{document}